\documentclass[%
 reprint,
 amsmath,amssymb,
 aps,
]{revtex4-2}

\usepackage{setspace}

\newcounter{inlineenum}
\renewcommand{\theinlineenum}{\roman{inlineenum}}
\newenvironment{inlineenum}
{\unskip\ignorespaces\setcounter{inlineenum}{0}%
	\renewcommand{\item}{\refstepcounter{inlineenum}{\textit{\theinlineenum})~}}}
{\ignorespacesafterend}

\usepackage{graphicx}
\usepackage{dcolumn}
\usepackage{bm}
\usepackage{siunitx} 

\begin{document}


    \title{Randomness in the choice of neighbours promotes cohesion in mobile animal groups}

\author{Vivek Jadhav}
 \email{vivekjadhav@iisc.ac.in}
\affiliation{Center for Ecological Sciences, Indian Institute of Science, Bengaluru, Karnataka, 560012, India}%
\author{Danny Raj M}%
 \email{dannym@iisc.ac.in}
\affiliation{%
Dept of Chemical Engineering, Indian Institute of Science, Bengaluru, Karnataka, 560012, India}%

\author{Vishwesha Guttal}%
 \email{guttal@iisc.ac.in}
\affiliation{%
Center for Ecological Sciences, Indian Institute of Science, Bengaluru, Karnataka, 560012, India}%

\date{\today}

\begin{abstract}
Classic computational models of collective motion suggest that simple local averaging rules can promote many observed group level patterns. Recent studies, however, suggest that rules simpler than local averaging may be at play in real organisms; for example, fish stochastically align towards only one randomly chosen neighbour and yet the schools are highly polarised. Here, we ask---how do organisms maintain group cohesion? Using a spatially-explicit model, inspired from empirical investigations, we show that group cohesion can be achieved even when organisms randomly choose only one neighbour to interact with. Cohesion is maintained even in the absence of local averaging that requires interactions with many neighbours. Furthermore, we show that choosing a neighbour randomly is a better way to achieve cohesion than interacting with just its closest neighbour. To understand how cohesion emerges from these random pairwise interactions, we turn to a graph-theoretic analysis of the underlying dynamic interaction networks. We find that randomness in choosing a neighbour gives rise to well-connected networks that essentially cause the groups to stay cohesive. We compare our findings with the canonical averaging models (analogous to the Vicsek model). In summary, we argue that randomness in the choice of interacting neighbours plays a crucial role in collective motion. 
\end{abstract}

\maketitle


\section{\label{sec:level1} Introduction}
Organisms that live in social groups often exhibit collective motion, which is important for achieving functions like foraging, navigation, evasion from predation, etc. \cite{camazine2020book, sumpter2010book, parrish1999complexity, ioannou2012predatory-science}. To explain the highly-coordinated motion of such animal groups, classic models of collective motion assume that an agent moves along the average direction of motion of all neighbours that are within a short metric distance around it~\cite{vicsek1995novelphasetransition}. Subsequent models extend on these ideas to incorporate cohesion~\cite{couzin2002collectivejtb, hemelrijk2005frontaldensity, strombom2011atrmodeljtb}; they assume that agents also move towards an average direction determined by the location of all nearby individuals. Broadly, theory and computational studies predict that non-trivial group-level phenomena can emerge even when organisms follow such simple rules that depend on the states of their neighbours~\cite{czirok1999collectiveptprl, toner1995long-rageprl, jhawar2019derivingmesoscopic, reynolds1987flocks, huth1992simulationjtb, vicsek1995novelphasetransition, couzin2002collectivejtb, parrish2002selfbiobul, hildenbrandt2010birdmodel}. 




Surprisingly, recent empirical studies~\cite{calovi2018disentanglingplosbio, lei2020computationalplosbio, jhawar2020noisenaturephy, biancalani2014noiseprl, dyson2015onset-pre} show that organisms interact through rules that are likely much simpler than averaging information of several individuals; in fact, each organism may interact with just a single randomly chosen neighbour (termed stochastic pairwise interaction) to achieve high levels of group polarisation~\cite{jhawar2020noisenaturephy}. This order, counter-intuitively, can emerge from sampling biases in the interactions due to the finite size of the group. Consequently, once the group is in a polarised state, it continues to reside in that state for long~\cite{jhawar2020noise-rs-pt}. To maintain group polarisation, group cohesion is a must. However, the mechanisms that keep the group cohesive--in particular the role of stochastic decision making---are not as well explored. 

Traditionally, to explain group cohesion, computational models assume that organisms interact with all individuals within a fixed metric distance~\cite{couzin2002collectivejtb, chate2008modeling-vicsek-var, gregoire2004onsetcm-prl}. However, several empirical investigations \cite{ballerini2008interaction-pnas, camperi2012spatially-inter-focus, Gautrais2012DecipheringGroups} have shown that organisms, in fact, interact with a select few, referred to as topological neighbours, and are not strictly limited by metric distances (say, 7 nearest neighbours in the case of starling flocks \cite{ballerini2008interaction-pnas}). They argue that such topological interactions provide substantially better cohesion than metric distance-based rules. In some fish schools, interaction with the nearest one appears to be sufficient in producing the observed cohesion~\cite{herbert2011inferring-pnas}. In another species~\cite{lei2020computationalplosbio}, fish appear to choose the most influential one among their neighbours to maintain cohesion. In echolocating organisms like bats, it is challenging to detect the neighbours as the returning echoes are faint and are likely masked by the loud calls of their neighbours. Consequently, in large groups, bats may only detect one neighbour at a time~\cite{beleyur2019modeling-bats}; yet roosting bats manage to maintain cohesion even in large mobile groups. While we do expect species-specific behavioural rules at fine scales, one broad question arises at this point: How does group cohesion depend on the choice of neighbours? In this context, we note that real organisms' behaviours are likely stochastic. While computational models do include an element of randomness for the motion of organisms, they typically ignore this in the context of choice of interacting neighbours (but see \cite{bode2010-stocnoise-jtb, bode2011limited-interface}). Here, we reveal the surprising role of randomness in the choice of neighbours in maintaining group cohesion.

In this article, we investigate how cohesion emerges from stochastic attraction interactions using a spatially explicit agent-based model. We explore a class of interaction models that differ only in the way the organism chooses its neighbour to interact with---based on \textit{randomness} in choice of neighbours. To understand how local interactions lead to cohesion at the group level, we reconstruct the underlying interaction network and employ a graph-theoretic analysis to study the properties of the network in light of how it is linked to the group's ability to stay cohesive. We compare our findings with the canonical equivalents of the averaging interactions to explain how simpler interactions are sufficient in achieving cohesion.

\begin{figure*}
    \centering
    \includegraphics[width=\textwidth]{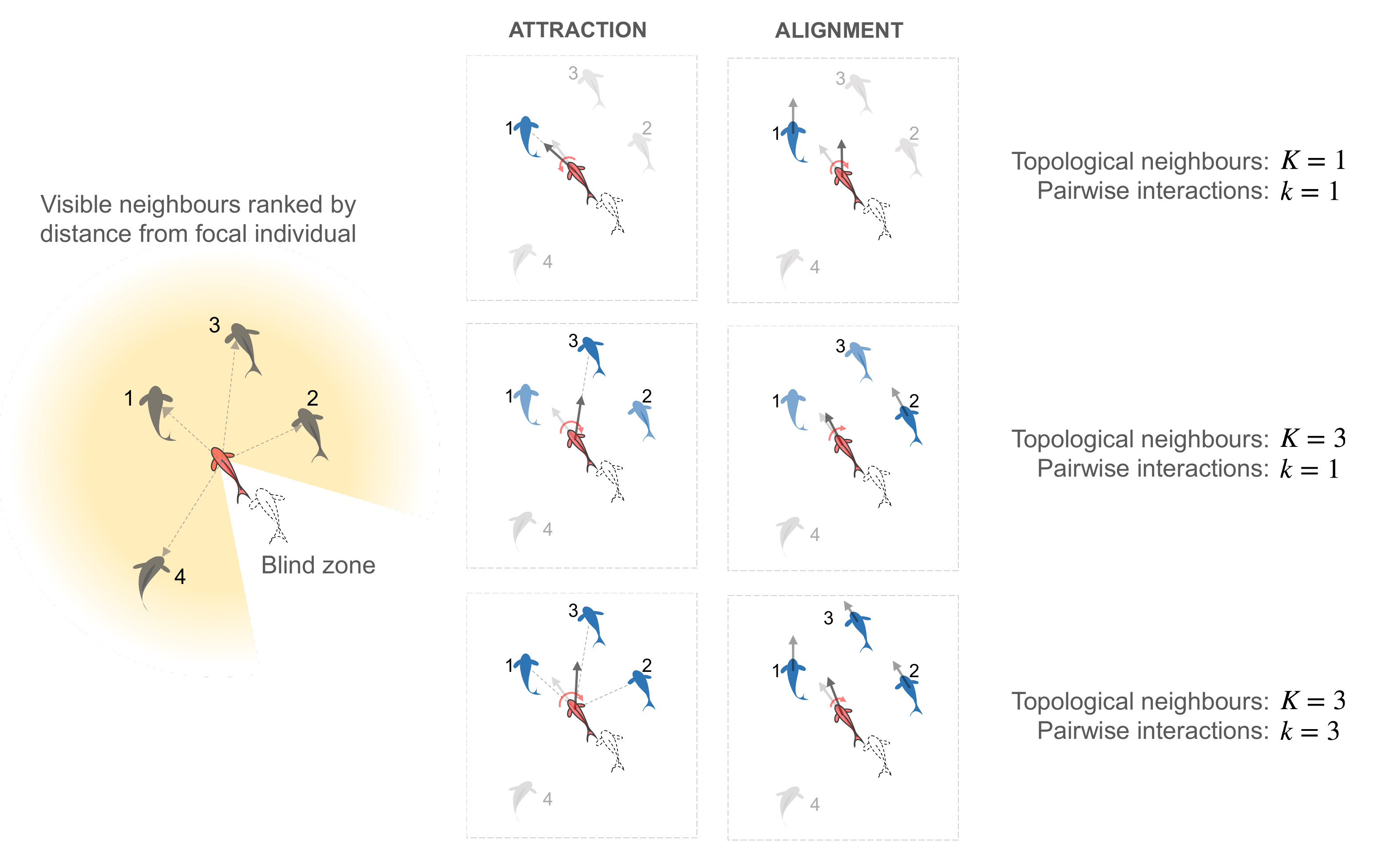}
    \caption{\textbf{Illustration of how organisms choose their neighbours to interact in the spatially-explicit agent-based model for mobile organisms (using schematics of fish schools).} A focal individual interacts with the neighbours it can visually perceive. The visible neighbours are ranked based on their distances from the focal individual. When $K=1$ and $k=1$, organisms either align with or move towards their nearest neighbour. When $K=3$ and $k=1$, organisms interact with one of their $3$ nearest neighbours. When $K=3$ and $k=3$, organisms interact with all $3$ of their nearest neighbours where they either align with the average orientation or move towards the centroid of the three organisms.}
    \label{fig:illustration_choosingK}
\end{figure*}

\section{\label{sec:Model} Model and simulation}

\subsection*{Agent-based spatially-explicit model}
We develop an agent-based model in two spatial dimensions to study the dynamics of collective motion. While our model broadly follows the principles of classic self-propelled particle models of collective motion \cite{Aoki1982simulation, huth1992simulationjtb, couzin2002collectivejtb}, we make some key modifications---such as probabilistic interaction rules, asynchronous updating and variable speed of agents---motivated by recent empirical studies \cite{katz2011inferring-pnas,herbert2011inferring-pnas,calovi2018disentanglingplosbio, jhawar2020noisenaturephy}. In this section, we only outline key model features and relegate the detailed mathematical descriptions to Appendix \ref{ax:fishmodeldetails}.

Every organism $i$ is described by its position in space ($\mathbf{x}_i$), and velocity (speed $s_i$ and direction $\mathbf{e}_i$). These are trivially related  by the kinematic equation: $\dot{\mathbf{x}_i} = s_i \mathbf{e}_i$. In our notation, while $\mathbf{e}_i$ represents the unit orientation vector, $\angle \mathbf{e}_i$ represents the angle of that orientation vector about the positive x-axis. 

Agents update both their speed and direction of motion as they interact with other agents in their visually perceptible neighbourhood. The stochastic decision making process of organisms is captured by the way the behavioural interactions of the organisms are implemented via asynchronous update rules and the choice of neighbours, as described below. 

Each organism asynchronously interacts to perform, \begin{inlineenum}
    \item exactly one behavioural interaction at a given instant (with a propensity that depends on the rate of the interactions),
    \item at a unique time (sampled from an exponential distribution).
\end{inlineenum}   

We assume that behavioural interactions between agents depend on {\it topological} neighbourhood \cite{ballerini2008interaction-pnas}: organisms perceive $K$ nearest neighbours from their visual zone. They integrate information---of orientation and position---from randomly chosen $k$ of the $K$ ($k \leq K$) perceived neighbours for both the alignment and attraction interactions (see Figure~\ref{fig:illustration_choosingK}). Our explorations show that choosing different neighbourhood sizes for alignment and attraction interactions do not alter our conclusions.

Four key behavioural rules are:

\paragraph{Alignment interaction:} At a rate $r_p$, an agent copies the speed and direction of motion of agents present in its topological neighbourhood. 

\paragraph{Attraction interaction:} At a rate $r_c$, an agent moves towards the centre of mass of its nearby topological neighbours with speed dependent on how far the organism is from this neighbourhood. 

 \paragraph{Spontaneous turning:} At a rate $r_s$, an agent chooses a new direction $\angle \mathbf{e}_d$, sampled from a (circular) normal distribution with a mean $\angle \mathbf{e}_i$ and a variance $\sigma_a^2$. The speed of the individual is sampled from a (truncated) normal distribution with mean $s_0$ and a variance $\sigma_s^2$. When the variances are small, spontaneous turning leads to a new direction of motion (with a new speed) but close to that of its previous direction. 
\paragraph{Collision avoidance:} Agents avoid collision with other agents as they move. We assume they do this by turning away from the other agent and by slowing down (as reported in \cite{herbert2011inferring-pnas}). 

\subsection*{Choice of neighbours and interaction types}
Depending on the values of $K$ and $k$, we can achieve a variety of interaction types. We primarily consider the following ones: 

We get a \textit{stochastic pairwise-interaction} when we set $k = 1$; e.g. relevant to fish~\cite{jhawar2020noisenaturephy, herbert2011inferring-pnas} and bats~\cite{beleyur2019modeling-bats} where organisms interact with only one randomly chosen neighbour. This results in the agent copying the direction of (alignment) or moving towards  (attraction), a single but randomly chosen neighbouring agent. Here, $K$ will determine the extent of the neighbourhood with which the organisms may interact.

Within the stochastic pairwise interaction models, when $K = 1$ an agent interacts with its nearest neighbour (referred to as \textit{nearest-neighbour-type}). At the other extreme, when $K=N-1$ an agent interacts with any agent from the entire group (where $N$ is the size of the group).

Moving beyond the stochastic pairwise interactions, we also consider the \textit{Vicsek-like local averaging} models by setting $k = K$ (with $K>1$).  For the alignment interaction, this corresponds to a topological analogue of the widely studied Vicsek interaction \cite{vicsek1995novelphasetransition}. For the attraction interactions, it mimics the topological analogue of the Boids~\cite{reynolds1987flocks} or the Couzin model \cite{couzin2002collectivejtb}. 

The simulation is carried out in a two-dimensional, unbounded (non-periodic) domain. Further details on initial conditions, parameter values, number of replicates, etc., are described in Appendix \ref{ax:fishmodeldetails}. More information on the sensitivity of the collective dynamics to parameter values can be found in Supplementary Information (SI) section 2.

Remarks comparing our model to some of the earlier models of collective motion are pertinent here. Most of the models assume that individuals interact with every other individual within a fixed metric distance. Further, they assume that all individuals interact and update their locations synchronously, moving at a constant speed \cite{vicsek1995novelphasetransition, couzin2002collectivejtb, lopez2012-review}. Following many recent empirical studies, our model assumptions differ on these aspects. First, we assume a topological distance for interactions. However, unlike Ballerini et al~\cite{ballerini2008interaction-pnas} who propose that birds interact with a fixed number of nearest neighbours, we assume that organisms may randomly choose any $k$ of the $K$ ($\ge k$) nearest neighbours; we show this randomness plays an important role in maintaining group cohesion. We also assume that individuals move at variable speeds \cite{hemelrijk2008self-var-speed, mishra2012collective-variable-speed, calovi2018disentanglingplosbio, Klamser_variablespeed} and update their motion asynchronously ~\cite{bode2010-stocnoise-jtb, bode2011limited-interface, strombom2019asynchrony}, in line with many recent empirical studies~\cite{katz2011inferring-pnas,herbert2011inferring-pnas,calovi2018disentanglingplosbio,jhawar2020noisenaturephy}. 
Note that the model we present in this section is a non-trivial extension of the mean-field model discussed in a previous work from our group~\cite{jhawar2020noisenaturephy}, which captures the polarisation order-parameter fluctuations observed in real fish. 

\paragraph*{Variants:} In order to show that the results obtained with this model are generic in nature and would arise in other contexts as well, we test the validity of our results with variants of the current model. We relax a key assumption in our model, asynchronicity, in two different ways to get two variants: \begin{inlineenum}
    \item Each agent interacts via all interactions simultaneously; however, different agents update their velocity at different times.
    \item All agents interact simultaneously and update their velocity at the same time.
\end{inlineenum} (more information in section 2 of SI).

\subsection*{Group cohesivity and quantification}
We first provide an intuitive picture of the group dynamics. When an agent is all by itself, the only interaction it undergoes is spontaneous turning, which changes its movement direction and speed at discrete points in time, resulting in random motion. When two or more agents are present, they begin to interact via alignment and attraction rules. In such a setup, the agents in the front often behave as though they were isolated since there is a visual field that limits the organism's perception. These agents have a greater tendency to wander away from the group until a spontaneous turn changes its orientation, allowing it to `see' a neighbour in the group; this may create the possibility for a successful attraction interaction at a later time that can bring it back to the group. At the same time, other group members may follow a wandering agent. When a few such individuals succeed in doing so, it could cause a cascade of many members leaving and eventually breaking a group into two or more smaller clusters. Thus, we investigate the interaction rules that maintain group cohesion. Before we do so, we need to define a way to quantify the cohesivity of groups. 

Cohesivity of a mobile group can be quantified in a number of different ways that include characterising the average near-neighbour distance \cite{ward2018-mixed-species-cohesion}, area of the convex hull of the spatial positions of the organisms and average distance to the barycenter of the group \cite{lei2020computationalplosbio}.
However, we choose a more stringent measure for cohesion based on how organisms are positioned in space with respect to their neighbours. Using a density-based spatial clustering approach (DBSCAN)\cite{ester1996dbscan, dbscan-matlab, ioannou2012predatory-science, guttal2010social-pnas}, we group the organisms into different clusters. 
Here, a cluster is defined as the set of all organisms, where every organism in the set is \textit{less than} $\epsilon$ distance away from at least one other organism in the set.
Then, we define a cohesion parameter (denoted as $\mathcal{C}$) as the average size (in number of organisms) of the largest cluster in the group normalised by the total size of the group (averaged over time of the simulation and over multiple realisations).
Hence, the value of $\mathcal{C}$ denotes the average fraction of organisms in the largest cluster in the group. A value of $1$ means that the organisms were always cohesive, with no break-up events.

\begin{figure*}
    \centering
    \includegraphics[width=\textwidth]{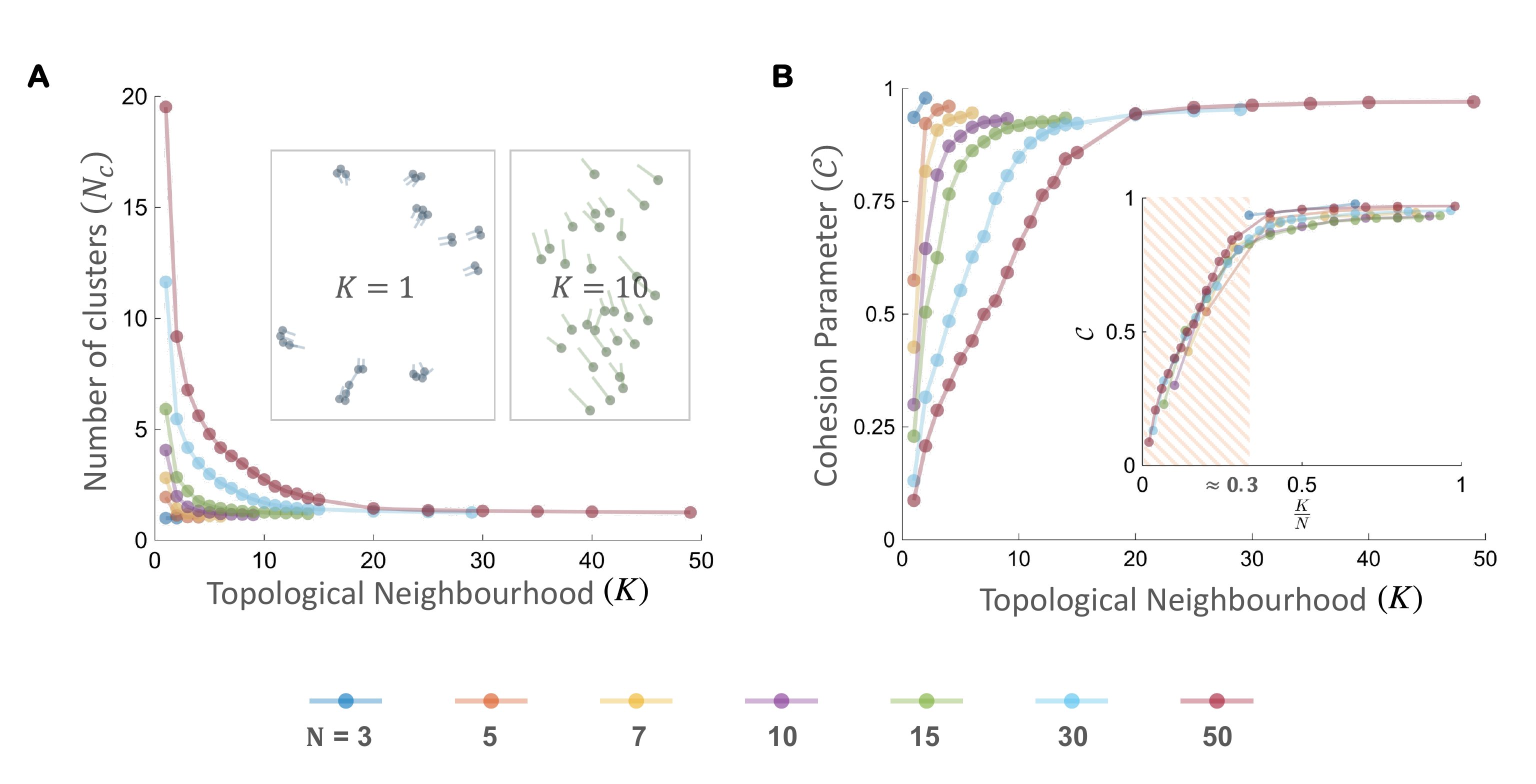}
    \caption{\textbf{High levels of cohesion are achieved in mobile groups even when organisms interact with just one neighbour (pairwise interaction) randomly chosen from a nearby neighbourhood of size $K$.} (A) Number of clusters $N_c$ a group breaks into, as a function of the neighbourhood of the organism $K$, for different group sizes. As $K$ increases, $N_c \to 1$. Inset: snapshots from simulation for size $N=30$ and $K=\{1, 10\}$. (B) Cohesion parameter $\mathcal{C}$ increases with neighbourhood size (K); Inset shows that cohesion reaches the maximum value when the neighbourhood is around 30\% of the group size across groups of several sizes.  The hashed region is $\frac{K}{N}<0.3$ where the change in $\mathcal{C}$ is significant.}
    \label{fig:cohesionresults}
\end{figure*}

\subsection*{Simulations and sensitivity analysis}
We simulate mobile groups of size ranging from $N\in[5, 50]$ and investigate the effect of the size of the neighbourhood an organism interacts with ($K\in{1,...,N-1}$). We first investigate the stochastic-pairwise interactions, where an organism interacts with only one randomly chosen neighbour, i.e. $k=1$ and then compare the findings with the canonical averaging interactions, by setting $k=K$.

The parameters used for the study are motivated by the empirical results from Jhawar et al \cite{jhawar2020noisenaturephy}. We analyse how sensitive our findings are to these parameters and metrics chosen for the study. We find that the qualitative features of group cohesion and its dependence on $K$ are insensitive to, \begin{inlineenum}
    \item a broad range of the parameters used in the spatially explicit agent based model,
    \item the parameters in calculating $\mathcal{C}$ and alternate definitions for the cohesion parameter, and
    \item the several variants of the spatial model.
\end{inlineenum} 
More information on this can be found in SI (Section 2).

\section{\label{sec:level3} Results and discussion}
\subsection*{Random choice of interacting neighbour, even with one individual, promotes group cohesion}
When the number of individuals are small, which correspond to group sizes in many simple experiments of collective motion~\cite{katz2011inferring-pnas,herbert2011inferring-pnas,calovi2018disentanglingplosbio}, individuals form groups and remain reasonably cohesive even when the interaction is of the near-neighbour-type, i.e. organisms align and attract with only the nearest neighbour ($K=1, k=1$; see Figure~\ref{fig:cohesionresults}A for $N=3, 5$). However, as the size of the group increases, interacting with just the nearest neighbour is no longer sufficient. Groups begin to break into smaller clusters of size $2$ or $3$ where the interactions between the organisms are confined to within the cluster. These clusters eventually drift away (see inset corresponding to $K=1$ of Figure \ref{fig:cohesionresults} A).

With increasing size of the topological neighbourhood while still interacting with only one random neighbour, i.e. larger $K$ but still with $k=1$, cohesivity of the groups ($\mathcal{C}$) increases, taking values close to $1$. This indicates that organisms reside in one tightly knit cluster stably throughout all time (see Figure \ref{fig:cohesionresults} B). This is likely because the number of unique individuals the organisms interact with increases with $K$. In contrast, when organisms interact only with their nearest neighbour, it is likely that they are interacting with the same neighbour repeatedly (for more information, see SI Section 3). 

We find the number of neighbours $K$ required to achieve the same level of cohesion scale with group size $N$. Simply put, the proportion of the topological neighbours required to achieve a given level of group cohesion is independent of the group size $N$. We find the cohesion parameter to saturate when the ratio of the topological neighbourhood to the total group size reaches a threshold of approximately $0.3$ (hashed region in Figure \ref{fig:cohesionresults} B). We find that this threshold ratio reduces when organisms' speed ($s_0$) reduces, or with increasing rates of attraction ($r_a$) and alignment ($r_p$) (see SI section 2).


\begin{figure*}
    \centering
    \includegraphics[width=\textwidth]{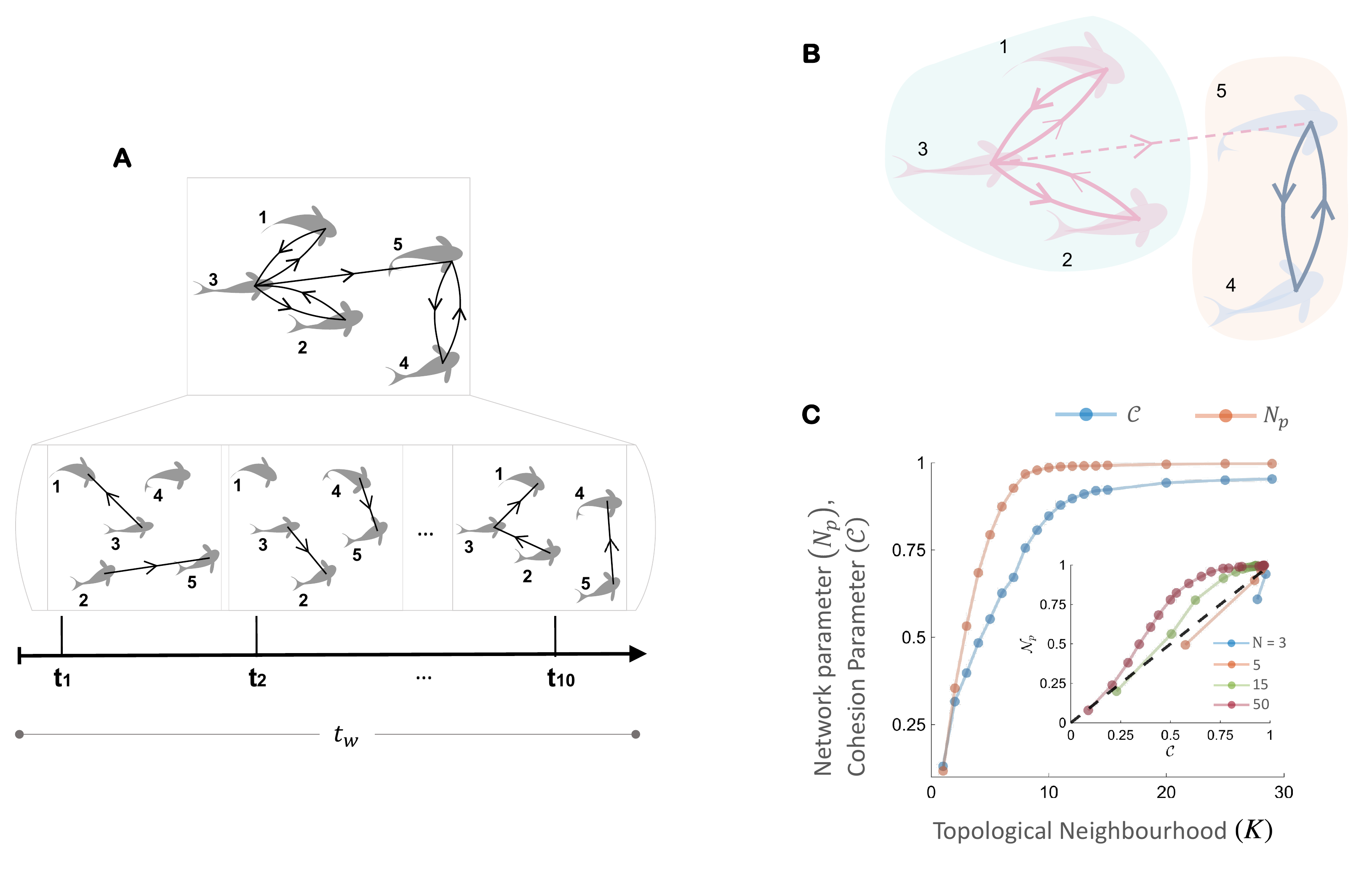}
    \caption{\textbf{Analysing the interaction network helps in understanding the emergence of cohesion in mobile groups (illustrated with the schematics of fish schools).} (A) Schematic outlining how an attraction interaction network is constructed, over a time window of $t_w$, with the knowledge of when an interaction occurred, between which of the organisms and at what time. (B) The interaction network is analysed to identify the sub-groups based on the connectivity $\tilde{\mathcal{A}}$. In the example shown, there are two sub-groups (1-3-2) and (4-5) (marked with different colours). The dotted line extending between the two sub-groups only connects them one way. (C) $\mathcal{C}$ and $\mathbf{N}_p$ are plotted as a function of $K$ for the group of size $N=30$. Inset: $\mathbf{N}_p$ is plotted as function of $\mathbf{C}$ for different group sizes. The Black dotted line marks the diagonal.}
    \label{fig:networkanalysis}
\end{figure*}

\subsection*{Attraction interaction network reveals why cohesion emerges}
To understand how cohesion emerges even from simple stochastic pairwise interactions, we turn to a graph-theoretic analysis of the underlying {\it attraction interaction network} between organisms in a group. We emphasise that we focus on attraction interactions rather than alignment interactions since our study is centred around how organisms maintain cohesion. While it is true that local directional alignment alone can also cause some degree of attraction, a major determining cause of group cohesion is the local attraction (see SI Section 4).

In our analysis, individual organisms can be considered as nodes, and a directed edge can be constructed from organisms $i$ to $j$, whenever $i$ exhibits an attraction interaction towards $j$. Since organisms interact asynchronously in our model, these edges are formed at distinct instants of time. Hence, to construct a graph that faithfully represents the underlying interaction network, we observe the different connections that arise between individuals over a time window $t_w$. To choose an appropriate time scale $t_w$ for the analysis, we use the length and velocity scales in the system corresponding to the motion of organism required to break free from its associated cluster: $\epsilon$ (maximum distance between organisms belonging to the same cluster) and $s_0$ (desired speed of an individual). The time scale is then defined as,
\begin{equation}\label{eqn:timescale}
    t_w = \frac{\epsilon}{s_0}
\end{equation}
Notice that if $t_w \gg 1$, then we would (at least in some cases) expect a network that is dense or fully connected since each organism would have interacted many numbers of times and if $t_w \ll 1$, the network would be sparse. Both these extremes would not represent the `correct' interaction network responsible for cohesion in a mobile group.

Figure \ref{fig:networkanalysis} A, illustrates how the attraction interaction network is constructed over a time period $t_w$. The network that emerges due to the interactions is directed in nature; i.e. $i \to j$ does not imply $j\to i$ since each individual randomly and asynchronously chooses a neighbour to interact. We argue that this underlying directed--network encodes information pertaining to the group's cohesiveness. Although there are studies that investigate network properties in collective motion models~\cite{aldana2003pt-noneqi-network,aldana2007-network-pt-prl, turgut2020interaction}, as far as we know, there are no off-the-shelf measures to characterise the interaction network to probe into why a group stays cohesive or breaks apart. 
In this section, we explore the correspondence between the properties of the network and the emergence of cohesion.

It is reasonable to expect a well-connected network to represent a cohesive group and a sparsely-connected one to represent a non-cohesive group. 
The simplest measure that characterises how well a network is connected is an adjacency matrix $\mathcal{A}$, where each element $(i,j)$ takes the value $1$ when there is an edge connecting $i$ to $j$. Note here that the directed nature of the graph results in a $\mathcal{A}$ that is asymmetric. 
From $\mathcal{A}$, the average connections for a node can be computed---which is simply the average number of neighbours an organism interacts with within time $t_w$. 
However, for cohesion to emerge, organisms need not necessarily interact with every other organism in the group. An organism interacting with just a few of its immediate neighbours could result in a chain of events that lead to cohesion.
To include this feature that arises not just from primary (or immediate) but from connections that are secondary, tertiary, etc., we compute the reachability matrix $\tilde{\mathcal{A}}$, where an element takes the value $1$ when there is a path from $i$ to $j$, in the directed graph.
To connect this property of the network to the group cohesivity, we divide a group into sub-groups based on $\tilde{\mathcal{A}}$.
A sub-group, in this context, is defined as a set of all organisms that have a path from and to every other organism in that sub-group. Then, we compute a network parameter $\mathcal{N}_p$ that is the size of the largest sub-group (normalised by the size of the group), averaged over time and several realisations (see Figure \ref{fig:networkanalysis} B for illustration; see Appendix \ref{ax:networkanalysisdetails} for details on numerical computation).

We find that $\mathcal{N}_p$ increases with the neighbourhood size $K$, in a manner qualitatively similar to the cohesion parameter $\mathcal{C}$ (see Figure \ref{fig:networkanalysis} C; also see SI section 5, where the network parameter is shown to describe the qualitative trends in $\mathcal{C} \text{ vs } K$ consistently for different levels of group cohesion).
When $K$ increases, interactions between organisms result in a network that is well-connected, \textit{i.e.} there is a path from every organism to almost every other in the sub-group, even when $K \simeq 0.3\times N$. This informs us that when organisms select individuals to interact with at random from a considerable topological neighbourhood, an opportunity is created for the group to stay cohesive. 

However, an interaction network created need not always materialise into a cohesive group. An organism can, in principle, interact with another organism in a cluster far away (in space) to create a well-connected network since the interactions in our model are topological. But other interactions like spontaneous turning, alignment or collisions, can break the network before it can cause the two clusters to come together. For this reason, we find $\mathcal{N}_p$ to reach a high value ($\simeq1$) faster than $\mathcal{C}$ for most cases (points over the diagonal in the inset of Figure \ref{fig:networkanalysis} C).
However, when group sizes are small ($N \le 5$) or when organisms break into small clusters (for the case of $K=1$), we find $\mathcal{N}_p$ to be lower than $\mathcal{C}$ (points under the diagonal in the inset of Figure \ref{fig:networkanalysis} C). These points refer to cases where the organisms are cohesive, but interactions are sparse, giving rise to a not fully connected network.
Here, a considerable number of organisms reside in the periphery of the clusters that do not have visible neighbours to interact with and hence get isolated from the rest of their neighbours in the calculation of the network parameter. Hence, these clusters have a lower value for $\mathcal{N}_p$ even though they are spatially in proximity to their local neighbours (SI Section 5).

\subsection*{Cohesion due to averaging interactions}
In canonical models for flocking, like the Vicsek model for alignment, an agent often averages the information from a neighbourhood to find its direction of movement.
Here we compare the group cohesivity achieved through stochastic pairwise interactions with that of the averaging-type interactions. We recall that while stochastic pairwise interactions are achieved by setting $k=1$ and $K>1$, we obtain the topological averaging interaction (like the Vicsek model for alignment) by setting $k=K$.

\begin{figure}
    \centering
    \includegraphics[width=\columnwidth]{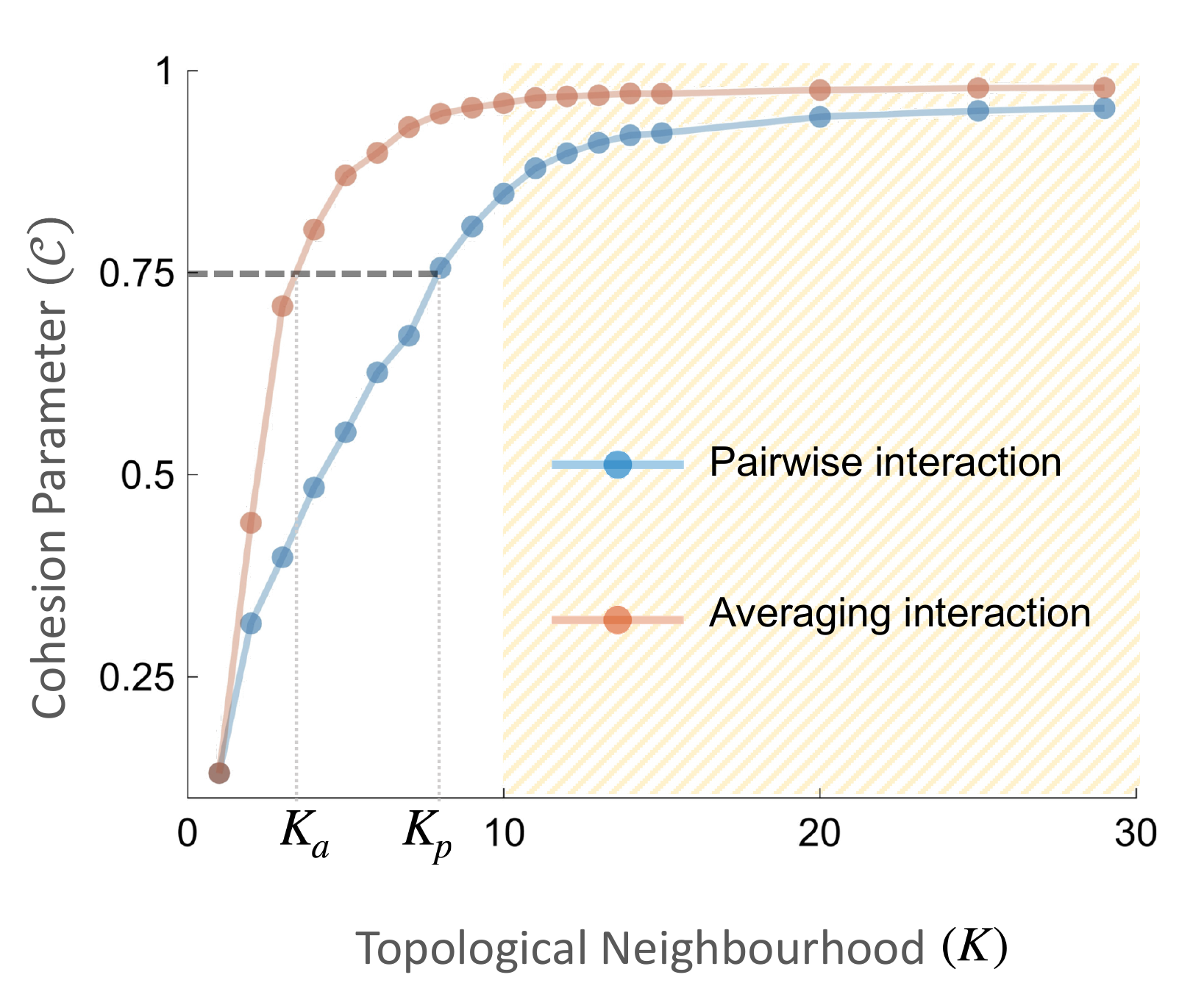}
    \caption{\textbf{Emergence of cohesion due to pairwise interactions ($k=1$) compared to averaging-type interactions ($k=K$) for a group of size $N=30$.} When $K$ is large, both interactions lead to cohesive mobile groups: hashed region. To achieve the same level of cohesion, a group interacting pairwise should have $K = K_p$ neighbours while a group interacting via averaging-interactions need only $K = K_a < K_p$ neighbours (horizontal line at $\mathcal{C}=0.75$).}
    \label{fig:PairwisevsAveraging}
\end{figure}

We find averaging interactions also achieve cohesion, with cohesion parameter $\mathcal{C}$ increasing rapidly with the size of the neighbourhood, $K$ (see Figure \ref{fig:PairwisevsAveraging}). We emphasise that while focal agents interact with all neighbours $K$ in the averaging-type interactions (because $k=K$), stochastic pairwise interactions permit interaction with only one neighbour ($k=1$). Averaging interaction, by definition, consumes information from all the interacting $K$ neighbours, while a pairwise interaction only takes-in information from only one of its $K$ neighbours at a time. Thus, it is not surprising that cohesion is achieved more rapidly in the local-averaging type interaction.  

Interestingly, beyond a certain value of neighbourhood size $K$, both the averaging and the pairwise interactions produce similar (maximum) cohesion. Hence, organisms interacting via these two different interaction types will not have any additional advantage with regards to cohesivity.
However, if we compared the neighbourhood sizes required to achieve a given value of cohesion (say, $0.75$), then we observe that the averaging interaction can achieve that level of cohesion with less number of neighbours $K_{a}$ than a pairwise interaction $K_{p}$ (see horizontal line at $\mathcal{C} = 0.75$ in Figure \ref{fig:PairwisevsAveraging}). 
From the viewpoint of the organism's cognitive capacity, the choice is between:
\begin{inlineenum}
    \item \label{list:cogcap-1} assimilating information from all neighbours in a small neighbourhood of size $K_{a}$ and averaging them or
    \item \label{list:cogcap-2} assimilating information of one neighbour from a larger set of $K_{p}$ neighbours.
\end{inlineenum}
While it is known that the cognitive load required to track a large number of neighbours is high ~\cite{lemasson2009collective-selective-attention,lemasson2013motion-social-navig,calvao2014-neigh-select-plosone}; we do not yet know, which of these two processes have a smaller cognitive load. 

However, one could safely assume that an organism capable of integrating information from multiple sources and limited by its ability to observe only a small part of its neighbourhood would prefer method--\ref{list:cogcap-1} over method--\ref{list:cogcap-2}, while a different organism that finds integrating information together difficult would choose \ref{list:cogcap-2} to achieve the same level of cohesion.

\section{Concluding remarks}
In this study, using a spatially-explicit agent-based model, we show that group-level cohesion can emerge when organisms move towards just one other randomly chosen nearby organism. We show that a random choice of the neighbour, rather than a fixed neighbour such as the nearest individual, considerably improves the group cohesion. Cohesion emerges even with such simple stochastic pairwise interactions because choosing a neighbour randomly creates a well-connected long-ranged interaction network. We show that the connectedness of the interaction network correlates well with the cohesivity of the mobile group. 

Constructing the interaction network was possible because we had complete access to all information pertaining to the interactions, their time-stamps and organisms-indices, due to the theoretical nature of the work. In an experimental setting, it would be challenging to estimate the underlying network structure from data of organismal motion.
In a recent study, a ray casting approach was used to identify a network based on the vision of individual fish \cite{rosenthal2015-hidden-network-pnas}. This network had an edge connecting an organism to every other organism in its perceivable neighbourhood; not specific to attraction interactions (or any other).
We believe that re-constructing the hidden interaction networks from movement data would be an exciting future direction for research.

\paragraph*{However, do organisms really choose a random neighbour to interact with?}
A random neighbour could be chosen in many ways. 
For instance, an organism could prefer a faster-moving individual to interact with, over other slower-moving ones. And since, organisms may move at different speeds during the course of their motion, which change continuously due to spontaneous activity and collisions, this `faster-individual' may be found anywhere within a neighbourhood of a certain size.
Lei and coworkers argue that fish choose to interact with a few of their most-influential neighbours \cite{lei2020computationalplosbio}. However, since the `influence' a neighbour has on a fish is a function of its proximity, relative positions and orientations, which change continuously as fish move in a school \cite{calovi2018disentanglingplosbio}, the most-influential fish could essentially take any position within the school at a given time: from the nearest neighbour to the farthest one. We speculate that choosing the most influential neighbour could be similar to choosing a fish randomly from a neighbourhood of size $K$. 

In summary, our study shows that when an organism randomly chooses another to interact with, irrespective of specific mechanisms, it results in an interaction network that is well-connected, giving rise to considerable group cohesion in small to intermediate group sizes.
However, we expect large systems of collective motion to exhibit dynamic fission and fusion; since the topological neighbourhood an organism should perceive to maintain group cohesion will be much larger than what is biologically feasible. 
We welcome further research on empirically motivated and parameterised models of collective motion that account for stochastic decision making of organisms with an emphasis on group cohesion, fission-fusion group dynamics and explore the functional significance of the role of heterogeneity between individuals in the group. 

\section{Data and code availability}
All codes are available via an open access repository: https://github.com/tee-lab/cohesion-in-collective-motion.git

\section{Author contributions}
VJ, VG and DRM conceptualised the study. VJ and DRM designed model and conducted the analysis. DRM and VJ wrote the manuscript with significant inputs from VG. All authors contributed to discussions and interpretations of results. 

\section{Acknowledgements}
We thank Arshed Nabeel for his comments on the manuscript. VJ is supported by a MHRD fellowship. DRM is supported by DST INSPIRE faculty fellowship program. VG acknowledges DBT-IISc partnership program and infrastructure support from DST-FIST. 


\bibliography{library1.bib}

\appendix

\section{\label{ax:fishmodeldetails} Variable speed model for mobile groups}
\subsection{Agent based simulations}
We employ a two dimensional model to simulate the collective motion of a mobile group of organisms. Motion of an organism depends on its instantaneous velocity (given by its direction of heading and speed), which evolves in time as it undergoes various interactions or encounters any obstacle for motion.
Equations \ref{eq:fishturning} and \ref{eq:fishspeeding} describe how an organism's heading and speed evolve in time. 
\begin{equation}
    \tau_{\theta} \frac{d \angle \mathbf{e}_i}{dt} = \angle \mathbf{e}_{d,i}-\angle \mathbf{e}_i + \Delta \theta_{r,i}
    \label{eq:fishturning}
\end{equation}

\begin{equation}
    \tau \frac{ds_i}{dt} = s_{d,i} - s_i + \Delta s_{r,i}
    \label{eq:fishspeeding}
\end{equation}
Every time an organism interacts, it gets a new desired direction $\mathbf{e}_{d,i}$ of motion and a new desired speed $s_{d,i}$ and the organism tries to achieve its new desired velocity in a time scale corresponding to $\tau$. The value of $\tau$ is related to its body mass and ability to turn/move quickly.
Each organism aligns, attracts or turns spontaneously with rates $r_p$, $r_c$ and $r_s$ respectively.
The time of the interaction is chosen stochastically: sampled from an exponential distribution with mean corresponding to the rate of the interaction. 
$\Delta \theta_{r,i}$ and $\Delta s_{r,i}$ in Eqs \ref{eq:fishturning} and \ref{eq:fishspeeding} correspond to the changes in speed and direction of an organism due to the presence of an obstacle; which could be another organism or a boundary. 

\textit{Note:} an organism moves continuously in space and time, making continuous adjustments to its trajectory when approaching an obstacle. Changes to its desired velocity due to interaction events occur only at discrete points in time.

\subsection{Behavioural interactions of the organisms}
In our simulations, we assume that organisms undergo three different types of interactions: spontaneous turning, alignment and attraction. In what follows we describe the mathematical details of these interactions.

\paragraph{Spontaneous turning:}
As an organisms moves, it spontaneously turns at a mean rate $r_s$, independent of its own state and the states of its neighbours.
The angle they wish to turn to, is sampled from a (circular) normal distribution with mean zero and variance $\sigma_a$ as in Eq \ref{eq:spontaneousAngle}.
\begin{equation}
    \angle \mathbf{e}_{d,i} = \angle \mathbf{e}_i + \mathcal{N}(0,\sigma_a)
    \label{eq:spontaneousAngle}
\end{equation}
The speed of the individual is sampled from a normal distribution with mean $s_0$, which is the desired velocity for movement of the organism and a variance $\sigma_s$ which quantifies the spread of the speeds of the individuals as observed in typical empirical data \cite{aoki1980analysis, bode2010perceived-proceedings}.
\begin{equation}
    s_{d,i} = \mathcal{N}(s_0,\sigma_s)
\end{equation}


\paragraph{Alignment:}
Every organism $i$, at an average rate $r_p$, aligns with its neighbour $j$ by copying its direction of movement $\mathbf{e}_j$ and speed $s_j$, over a time scale $\tau$ (see eqs \ref{eq:alignmentAngle} and \ref{eq:alignmentSpeed}). 
\begin{equation}
    \angle \mathbf{e}_{d,i} = \angle \mathbf{e}_j
    \label{eq:alignmentAngle}
\end{equation}
\begin{equation}
    s_{d,i} = s_j
    \label{eq:alignmentSpeed}
\end{equation}
The Equations \ref{eq:alignmentAngle} and \ref{eq:alignmentSpeed} correspond to the case where the organism interacts with just one of its nearby neighbour.
For the general case, where an organism interacts with $k$ of its $K$ nearby, visible neighbours the interactions take the following form.
\begin{equation}
    \angle \mathbf{e}_{d,i} = \frac{1}{k}\sum_{j=1}^k \angle \mathbf{e}_j
    \label{eq:alignmentAngle_generic}
\end{equation}
\begin{equation}
    s_{d,i} = \frac{1}{k}\sum_{j=1}^k s_j
    \label{eq:alignmentSpeed_generic}
\end{equation}


\paragraph{Attraction:}
An organism $i$ moves towards another $l$, at an average rate, $r_c$, to stay cohesive. It is reasonable to expect $i$ to change its current trajectory to move towards $l$ only when it is considerably far away from $l$. This feature is brought into the interaction term by using a distance based weighting $f_{a,l}$, as shown in Eq \ref{eq:attractionAngle}.
\begin{equation}
    \angle \mathbf{e}_{d,i} = \angle \big[\mathbf{e}_i + f_{a,l} \mathbf{r}_{il} \big]
    \label{eq:attractionAngle}
\end{equation}
We can also model the change in speed experienced by $i$ in a similar fashion: when $i$ is far away from $k$, it moves faster (see eq \ref{eq:attractionSpeed}).
\begin{equation}
    s_{d,i} = s_{0} + \kappa_{a} f_{a,l}
    \label{eq:attractionSpeed}
\end{equation}
Again Eq \ref{eq:attractionAngle} and \ref{eq:attractionSpeed} correspond to pairwise interactions where $k=1$. For the generic case, the interactions take the following form.
\begin{equation}
    \angle \mathbf{e}_{d,i} = \angle \big[\mathbf{e}_i + \frac{1}{k}\sum_{l=1}^k  f_{a,l} \mathbf{r}_{il} \big]
    \label{eq:attractionAngle_generic}
\end{equation}
\begin{equation}
    s_{d,i} = s_{0} + \kappa_{a} \frac{1}{k}\sum_{l=1}^k f_{a,l}
    \label{eq:attractionSpeed_generic}
\end{equation}
The weighting function selected for use in our model has the form shown in eq \ref{eq:attractionWeighting}. When $\gamma$ is $>2$, it softens the effect of the attractive interaction within distance $l$ around the organism $l$. Only when $(\|\mathbf{r}_{ik}\| - 2 R) > l$, the attractive interaction has a significant effect.
\begin{equation}
    f_{a,l} = \Bigg(\frac{\|\mathbf{r}_{il}\| - 2 R}{l}\Bigg)^\gamma
    \label{eq:attractionWeighting}
\end{equation}

\paragraph{Collision avoidance:}
Organisms avoid collision with each other during collective motion.
They turn towards a direction that is perpendicular to that connecting the organisms as shown in eq \ref{eq:collavoidanceAngle}, to prevent collision events. 
\begin{equation}
    \Delta \theta_r = \angle \Bigg[\frac{1}{n_r} \sum_{o=1}^{n_r} \lambda_a \lambda_p \big[r_{io, \perp}\big]\Bigg]
    \label{eq:collavoidanceAngle}
\end{equation}
Here, $n_r$ is the number of individuals that are within the zone of repulsion $(zor)$: which is the critical distance between organisms below which they begin to respond to the proximity of their neighbours.
In addition to turning, they also slow down as they approach an obstacle, as shown in eq \ref{eq:collavoidanceSpeed}. The rate of slowing down is inversely proportional to the distance between the fish and directly proportional to the rate of approach.
\begin{equation}
    \Delta s_r =  \frac{\kappa_{r}}{n_r} \sum_{o=1}^{n_r} \lambda_a \lambda_p \Bigg[ \frac{s_i\mathbf{e}_i.\mathbf{r}_{io} +s_o\mathbf{e}_o.\mathbf{r}_{oi}} {\big(\|\mathbf{r}_{io}\| - 2 R\big)^\beta}\Bigg]
    \label{eq:collavoidanceSpeed}
\end{equation}

The parameters $\lambda_a$ and $\lambda_p$ are binary variables that take the value $1$ only when the fish $i$ moves towards fish $o$ and the distance between the two fish is decreasing in time. They are defined using a Heaviside functions as shown in eqs \ref{eq:parametersCollAvoidanceprox} and \ref{eq:parametersCollAvoidanceapproach}. 
\begin{eqnarray}
    \label{eq:parametersCollAvoidanceprox}
    \lambda_a = \mathcal{H}[e_i.r_{io}]\\
    \label{eq:parametersCollAvoidanceapproach}
    \lambda_p = \mathcal{H}[e_i.r_{io} + e_o.r_{oi}]
\end{eqnarray}

\paragraph{Remark: bounds on speed:}
The speed $s_i$ of an organism, is bound between $0$ and $s_{max}$, which is set to $10 \si{cm s^{-1}}$ in our study. This is enforced in both the spontaneous turning and attraction interactions.

\paragraph{Remark: local state-dependent behaviour:}
We model organismal--movement using agents that interact with each other at a rate, intrinsic of the agent, \textit{i.e.} independent of the local state of the group. Simply put, a fish aligns with a neighbouring fish, irrespective of whether the local neighbourhood is ordered or not. Similarly, a fish does not move towards its neighbour in response to its perception of the less cohesive neighbourhood.
However, interestingly, the implementation of the alignment and attraction rules, bring an indirect local-state dependent behaviour. 
An organism aligning with a neighbour does not change the local order significantly when the local neighbourhood is sufficiently ordered. While, this change is significant when the neighbourhood is disordered. The weighting function $f_{a,l}$ as described in eq \ref{eq:attractionWeighting}, makes sure that the effect of the attraction interaction is significant only when the organism is away from the local neighbour it is interacting with.
Even though organisms interact in a manner, independent of the local state of the system, the interaction events themselves cause significant change in the local group-properties only when the local state is away from that the fish desires to achieve.

\paragraph{Numerical simulations:}
The codes for the agent based model were developed in-house and the numerical simulations are carried out in MATLAB \copyright. We ran simulations for T = 3500s with integration time of dt = 0.05s for 24 realisations. Codes have been made available (see code availability).

Model parameter values used for the results in the main text are given in (Table~\ref{para-table}).\\

\setlength{\arrayrulewidth}{0.5mm}
\setlength{\tabcolsep}{12pt}
\renewcommand{\arraystretch}{1.5}
\begin{singlespace} 
\begin{table}
\centering
\begin{tabular}{p{5cm} | p{2cm}| p{1.2cm} | p{1.2cm}}
\hline
 Parameter & Unit & Symbol & Value\\
 \hline
 Spontaneous rate & \si{s^{-1}} & $r_{s}$ & 1 \\
 Alignment rate & \si{s^{-1}} & $r_{p}$ & 1.5 \\
 Attraction rate & \si{s^{-1}}& $r_{a}$ & 1 \\
 Visual field &  Degree & $\theta_{s}$ & $270^{0}$\\
 Desired Speed & \si{cm ~s^{-1}} & $s_{0}$ & 0.2\\
 Maximum Speed & \si{cm ~s^{-1}} & $s_{max}$ & $5\times s_{0}$\\
 Agent size & cm & $R$ & 0.2\\
 Variance of angular displacement (spontaneous turn) & $\text{rad}^{2}$ & $\sigma_{a}$ & $\frac{\theta_{s}\times \pi}{2\times 180}$\\
 Variance of speed (spontaneous turn) & \si{cm^{2} ~s^{-2}} & $\sigma_{s}$ & $v_{0}$\\
 Relaxation time for speed & s & $\tau$ & 0.2\\
 Relaxation time for angular speed & s & $\tau_{\theta}$ & 0.5 \\
 Distance-dependent attraction (coefficient) & \si{cm ~s^{-1}} & $\kappa_{a}$ & $ 10^{-2}$ \\
 Distance-dependent attraction (order) & 1 & $\gamma$ & 3\\
 Zone of repulsion  & cm & $zor$ & 1.2\\
 Distance-dependent repulsion (coefficient) & \si{cm^{\beta}} & $\kappa_{r}$ & $-10^{-3}$\\
 Distance-dependent repulsion (order) & 1 & $\beta$ & 3\\
 Maximum distance between agents to belong to the same cluster & cm & $\epsilon$ & 2.4\\
\hline
\end{tabular}
\caption{Summary of model parameters.}
\label{para-table}
\end{table}
\end{singlespace}

\section{\label{ax:networkanalysisdetails} Analysing the interaction networks}

We begin by constructing an adjacency matrix $\mathcal{A}$ where $\mathcal{A}_{ij} = 1$ whenever an organism $i$ (represented by node $i$ in the graph) interacts with organism $j$ (node $j$) through an attraction interaction. 
This process is continued for all the similar interactions during the time window $t_w$.
Irrespective of how many times $i$ and $j$ interact, only a single edge is constructed between the two. 

We use this adjacency matrix $\mathcal{A}$ to construct a directed graph, $\mathcal{G}$. We use a MATLAB \cite{MATLAB:R2021b} function \verb|digraph| to construct the directed graph, $\mathcal{G}$. Once we construct $\mathcal{G}$, we use another function in MATLAB, \verb|conncomp| to analyse the connectivity of the graph. \verb|conncomp| groups the nodes based on their connectivity. Node $i$ and $j$ belong to the same group if there is path from $i$ to $j$ and from $j$ to $i$ \cite{cormen2009introduction}. We then define the network parameter $\mathcal{N}_{p}$ as the average size of the largest group divide by total number of nodes (organisms).

\end{document}